\documentclass[12pt]{iopart}
\usepackage{iopams}   
\usepackage[colorlinks=true,urlcolor=blue,citecolor=blue,linkcolor=blue]{hyperref}
\usepackage[T1]{fontenc}
\usepackage[latin9]{inputenc}
\usepackage{amssymb}
\usepackage{graphicx}
\usepackage{color}
\usepackage{mathrsfs}
\usepackage{float}
\usepackage{indentfirst}
\usepackage{color}

\usepackage{amsthm}
\usepackage{hyperref}
\usepackage{amscd}
\usepackage[english]{babel}

\newcommand{\diag}{\mathop{\mathrm{diag}}\nolimits}

\usepackage{tikz}
\usetikzlibrary{arrows}

\begin{document}

\title{Fermionic Quantum Critical Point of Spinless Fermions on a Honeycomb Lattice}

\author{Lei Wang$^{1}$, Philippe Corboz$^{1,2}$ and Matthias Troyer$^{1}$}

\address{$^{1}$Theoretische Physik, ETH Zurich, 8093 Zurich, Switzerland}
\address{$^{2}$Institute for Theoretical Physics, University of Amsterdam, Science Park 904 Postbus 94485, 1090 GL Amsterdam, The Netherlands}

\ead{lewang@phys.ethz.ch}

\begin{abstract}
Spinless fermions on a honeycomb lattice provide a minimal realization of lattice Dirac fermions. Repulsive interactions between nearest neighbors drive a quantum phase transition from a Dirac semimetal to a charge-density-wave state through a \emph{fermionic} quantum critical point, where the coupling of Ising order parameter to the Dirac fermions at low energy drastically affects the quantum critical behavior. Encouraged by a recently discovery~\cite{Huffman:2014fj} of absence of the fermion sign problem in this model, we study the \emph{fermionic} quantum critical point using the continuous time quantum Monte Carlo method with worm sampling technique. We estimate the transition point $V/t= 1.356(1)$ with the critical exponents $\nu =0.80(3)$ and $\eta =0.302(7)$. Compatible results for the transition point are also obtained with infinite projected entangled-pair states.
\end{abstract}



\maketitle

\section{Introduction}
Interaction induced quantum phase transitions of Dirac fermions are of general interests in graphene~\cite{CastroNeto:2009cl}, $d$-wave superconductors~\cite{SachdevBook}, topological insulators~\cite{Hasan:2010ku}, ultracold atoms~\cite{Uehlinger:2013bj} and high energy physics~\cite{Rosenstein:1991tj}. One of the prototypical examples consists of half-filled spinless fermions on a honeycomb lattice interacting through nearest neighbor repulsions 

\begin{eqnarray}
 \hat{H}& = & \hat{H}_{0} + \hat{H}_{1}, 
   \label{eq:Ham} \\
  \hat{H}_{0} & = & -t \sum_{\langle \mathbf{i,j} \rangle} \left( \hat{c}_{\mathbf{i}}^{\dagger} \hat{c}_{\mathbf{j}} + \hat{c}_{\mathbf{j}}^{\dagger} \hat{c}_{\mathbf{i}}\right)
   = \sum_\mathbf{i,j} \hat{c}^{\dagger}_{\mathbf{i}} K_\mathbf{ij} \hat{c}_{\mathbf{j}},  \label{eq:K} \\
  \hat{H}_{1} & = & V \sum_{\langle \mathbf{i,j} \rangle} \left( \hat{n}_{\mathbf{i}} - \frac{1}{2}
  \right) \left( \hat{n}_{\mathbf{j}} - \frac{1}{2} \right). 
  \label{eq:V}
\end{eqnarray} 
Eq.(\ref{eq:Ham}) is arguably the simplest model exhibiting a quantum phase transition of Dirac fermions in two dimension. However, despite its deceptively simple form, the model exhibits an \emph{unconventional} quantum critical point which deserves detailed study because it may lay the foundation of understanding rich phenomena when other degrees of freedom or intertwined phases are involved. 

\begin{figure}[t]
\centering
\includegraphics[width=9cm]{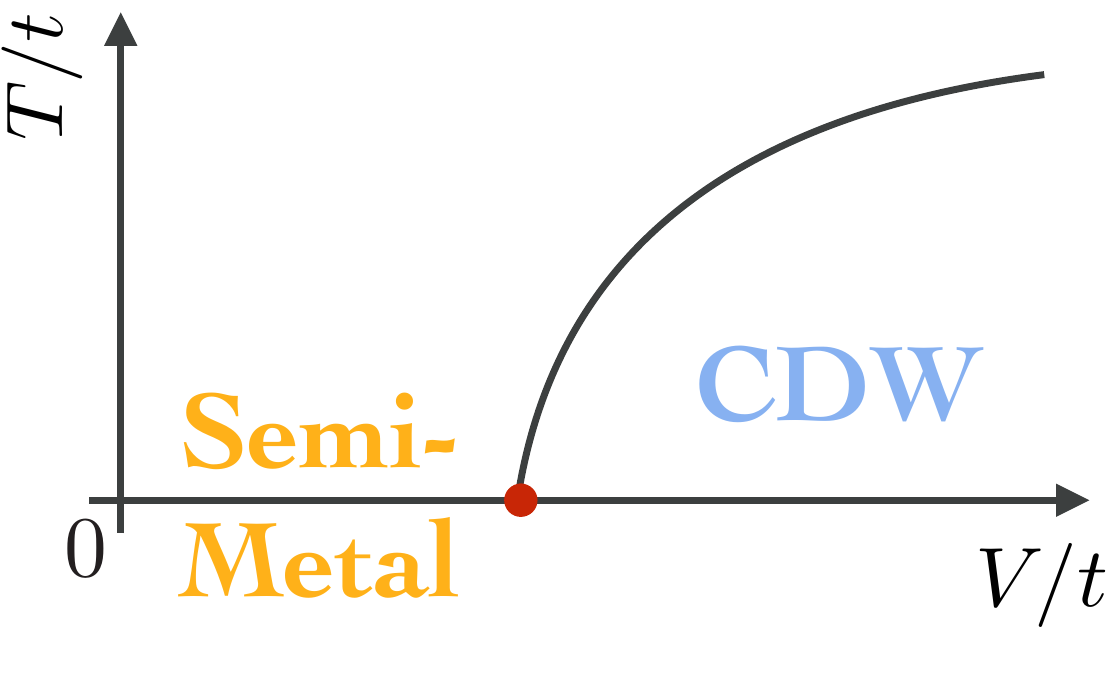}
\caption{Schematic phase diagram of model Eq.(\ref{eq:Ham}). In strong coupling limit the system is in the charge-density-wave state. The long range order melts through a 2D Ising transition upon increase of temperature or a quantum phase transition upon decrease of $V/t$. The red dot represents a \emph{fermonic} quantum critical point which is the focus of this paper. }
\label{fig:phasediag}
\end{figure}

The phase diagram of Eq.(\ref{eq:Ham}) is easy to anticipate, see Fig.\ref{fig:phasediag}. The system behaves like a classical lattice gas in the strong coupling limit ($V\gg t$) and favors a staggered charge-density-wave (CDW) state. The CDW state breaks the discrete sublattice symmetry and melts through a 2D Ising phase transition at finite temperature. In the weak coupling limit, quantum fluctuations due to fermion hopping destroy the CDW long range oder and restore the Dirac semimetal state. Since  Dirac fermions are perturbatively stable against short range interactions, the quantum critical point separating the Dirac semimetal and the CDW state lies at a finite interaction strength $V/t$. 

The topology of the phase diagram Fig.\ref{fig:phasediag} resembles that of the familiar 2D transverse field Ising model~\cite{SachdevBook} where quantum fluctuations induced by the transverse field destroy the Ising long range order. However, in model Eq.(\ref{eq:Ham}) the coupling of the Ising order parameter to the Dirac fermions at low energy strongly affects its quantum critical behavior. It cannot be treated by the familiar scalar $\phi^{4}$-theory since integrating out the Dirac fermions will lead to a singular action for the Ising fields. The low energy physics is described by the Gross-Neveu-Yukawa theory~\cite{Gross:1974uk, Herbut:2006jaa, Herbut:2009ga, Herbut:2009bb} which features a \emph{fermonic} quantum critical point. The Gross-Neveu-Yukawa theory has been studied intensively in the context of high energy physics~\cite{Rosenstein:1993wh, Rosa:2001en,Hofling:2002bl} and quantum critical point scenario of the $d$-wave superconductors~\cite{Laughlin:1998tp,Vojta:2000dd, Khveshchenko:2001kz}, however, there is no consensus concerning the critical exponents, partially due to uncontrolled approximations involved in various theoretic approaches. 

Quantum Monte Carlo (QMC) simulations are valuable unbiased approach to study the quantum critical behavior if the notorious fermion sign problem is absent~\cite{Troyer:2005hv}. A recent example is the study of Dirac semimetal to antiferromagnetic insulator transition in the half-filled repulsive Hubbard model on the honeycomb lattice~\cite{Sorella:1992wd, Meng:2010gc, Sorella:2012hia, Assaad:2013kg}. Unfortunately, the seemingly simpler model Eq.(\ref{eq:Ham}) has a severe sign problem in the conventional auxiliary field QMC method~\cite{Blankenbecler:1981vj}. Early QMC studies have thus been limited to high temperatures or small system sizes~\cite{Scalapino:1984wz,Gubernatis:1985wo}. The meron-cluster algorithm~\cite{Anonymous:xaWVK-gC} solves the sign problem for $V\ge 2t$ and simulations using it confirm the finite temperature Ising transition of several staggered fermion models~\cite{Anonymous:UJp124el, Anonymous:WZs4V-GU}. However, the quantum critical point of the model Eq.(\ref{eq:Ham}) lies at $V<2t$ and is not accessible by the meron-cluster algorithm. 
The Fermi bag approach~\cite{Chandrasekharan:2010gk} has been used to study the 3D lattice massless Thirring model~\cite{Chandrasekharan:2012br} and the Gross-Neveu model~\cite{Chandrasekharan:2013hta} with two flavors of four component Dirac fermions. 

Recently, Ref.~\cite{Huffman:2014fj} discovered that the sign problem of the model Eq.~(\ref{eq:Ham}) is absent in the continuous-time quantum Monte Carlo (CTQMC) formalism~\cite{Rubtsov:2005iw,Gull:2011jd}. This allows us to access the quantum critical point in the CTQMC simulation. Using a standard finite size scaling analysis we estimate the critical point $V/t\approx 1.356$ and the critical exponents $\nu\approx0.8,\eta \approx 0.3$. Our results are summarized in Table~\ref{tab:exponents}. We believe these results do not only apply to the specific microscopic model Eq.(\ref{eq:Ham}), but also hold for many intriguing problems including the chiral symmetry breaking of Dirac fermions~\cite{Rosenstein:1991tj} and the quantum critical point in the $d$-wave superconductors~\cite{SachdevBook}. Future theoretical or experimental advances in either field~\cite{CastroNeto:2009cl, SachdevBook, Hasan:2010ku, Uehlinger:2013bj,Rosenstein:1991tj} will be able to test our  predictions. 

The paper is organized as follows: In Sec.\ref{sec:ctqmc} we briefly review the CTQMC method with focuses on the absence of the sign problem~\cite{Huffman:2014fj}. In Sec.~\ref{sec:worm} we introduce the worm sampling technique in the Monte Carlo calculation~\cite{Prokofev:1998tc, Burovski:2006hv, Gull:2011jd} to improve the efficiency of the simulation. Section~\ref{sec:results} contains our main results and discussions as well as  comparisons with results obtained by infinite projected entangled-pair states (iPEPS) calculations. In Sec.\ref{sec:outlook} we briefly anticipate several future research directions based on this work. In the appendix we provide technique details of our numerical calculation (\ref{sec:updates}) and additional results obtained on the $\pi$-flux square lattice (\ref{sec:piflux}).


\section{Method}
Two properties of the model (\ref{eq:Ham}) are essential for the absence of the sign problem in the CTQMC simulation. First, the filling is fixed at $1/2$ per site because of the particle-hole symmetry of the model. Second, the hopping matrix defined in Eq.(\ref{eq:K}) satisfies

\begin{eqnarray}
  K_\mathbf{ji} & = & -\eta_\mathbf{i} K_\mathbf{ij} \eta_\mathbf{j}. 
\label{eq:KT}
\end{eqnarray}
where the ``parity index'' $\eta_\mathbf{i}=1(-1)$ for a site $\mathbf{i}\in A(B)$ sublattice. 

\subsection{Interaction expansion CTQMC and the sign problem \label{sec:ctqmc}}
We expand the partition function of the system in terms of the interaction vertices of Eq.(\ref{eq:V}) ~\cite{Rubtsov:2005iw,Gull:2011jd}
\begin{eqnarray}
  Z & = & Z_{0}\sum_{k=0}^{\infty} \frac{( -V )^{k}}{k!} \int_{0}^{\beta}\mathrm{d}  \tau_{1}\int_{0}^{\beta} \mathrm{d} \tau_{2} \ldots
  \int_{0}^{\beta} \mathrm{d} \tau_{2k} \,\delta(\tau_{1}-\tau_{2}) \ldots \delta(\tau_{2k-1}-\tau_{2k}) \times  \nonumber \\  && \left\langle \left(
  \hat{n}_{\mathbf{i}_{1}} ( \tau_{1} ) - \frac{1}{2} \right) \left( \hat{n}_{\mathbf{i}_{2}} (
  \tau_{2} ) - \frac{1}{2} \right) \ldots \left( \hat{n}_{\mathbf{i}_{2k-1}} ( \tau_{2k-1} )
  - \frac{1}{2} \right) \left( \hat{n}_{\mathbf{i}_{2k}} ( \tau_{2k} ) - \frac{1}{2}
  \right) \right\rangle_{0} \nonumber \\
   & = & Z_{0}\sum_{k=0}^{\infty} \frac{( -V )^{k}}{k!} \int_{0}^{\beta} \mathrm{d}  \tau_{2} \int_{0}^{\beta} \mathrm{d} \tau_{4}  \ldots
  \int_{0}^{\beta}  \mathrm{d} \tau_{2k} \,  \det ( G^{k}), 
\label{eq:Z}
\end{eqnarray}
where $ \hat{n}_{\mathbf{i}} ( \tau ) = e^{\hat{H}_{0}\tau} \hat{n}_{\mathbf{i}}  e^{-\hat{H}_{0}\tau} $ and $Z_{0}$ is the partition function of the noninteracting system. $\langle\ldots\rangle_{0}=\mathcal{T}\Tr (e^{-\beta \hat{H}_{0}}\ldots)/Z_{0}$ denotes the average over the noninteracting Hamiltonian Eq.(\ref{eq:K}) and $\mathcal{T}$ is the time ordering operator. The interaction vertices $\{\mathbf{i}_{1} ,\mathbf{i}_{2} \}, \ldots , \{\mathbf{i}_{2k-1} ,\mathbf{i}_{2k} \}$ consist of $k$ pairs of neighboring sites. The delta functions in the first line of Eq.(\ref{eq:Z}) indicates that the interactions are instantaneous. $G^{k}$ is a $2k\times 2k$ matrix 
\begin{eqnarray}
G^{k}_{pq} = \mathcal{G}^{0}_{\mathbf{i}_{p}\mathbf{i}_{q}}(\tau_{p}-\tau_{q})-\delta_{pq}/2,
\label{eq:G}
\end{eqnarray}
where $\mathcal{G}^{0}_\mathbf{ij}(\tau)= \langle \hat{c}_\mathbf{i} ( \tau ) \hat{c}_\mathbf{j}^{\dagger}\rangle_{0}$ is the noninteracting Green's function. The particle-hole symmetry ensures that $\mathcal{G}^{0}_{\mathbf{i} \mathbf{i}}(0^{+}) =1/2$ and therefore the diagonal element of $G^{k}$ vanishes. 
In addition, one has 
\begin{eqnarray}
  \mathcal{G}^{0}_{\mathbf{i}_{p} \mathbf{i}_{q}} ( \tau >0 )  
 = \left( \frac{e^{-K \tau}}{1+e^{- \beta K}} \right)_{\mathbf{i}_{p} \mathbf{i}_{q}}, 
  \label{eq:gpq}
 \end{eqnarray}
while using the anti-periodicity of the Green's function and Eq.(\ref{eq:KT}) one has 
 \begin{eqnarray}
  \mathcal{G}^{0}_{\mathbf{i}_{q} \mathbf{i}_{p}} ( - \tau <0 ) 
= - \left( \frac{e^{-K ( -\tau + \beta )}}{1+e^{- \beta K}} \right)_{\mathbf{i}_{q} \mathbf{i}_{p}}  
 =  -\mathcal{G}^{0}_{\mathbf{i}_{p} \mathbf{i}_{q}} ( \tau >0 ) \eta_{\mathbf{i}_{p}}\eta_{\mathbf{i}_{q}}. 
 \label{eq:gminustau}
\end{eqnarray} 
Equations (\ref{eq:G}-\ref{eq:gminustau}) show that the Green's function matrix satisfies $G^{k}_{qp} =  -\eta_{\mathbf{i}_{p}}  G^{k}_{pq}  \eta_{\mathbf{i}_{q}} $. Introducing a diagonal matrix $D^{k}  =  \diag(\eta_{\mathbf{i}_{1}},\eta_{\mathbf{i}_{2}}, \ldots, \eta_{\mathbf{i}_{2k}})$, it can be written as 
\begin{eqnarray}
 (G^{k} D^{k})^{T}  & =& - G^{k}D^{k}  
 \label{eq:skewsymm}
  \end{eqnarray}
In other word, the matrix $G^{k}D^{k}$ is skew-symmetric and its determinant is non-negative because it equals the square of the Pfaffian of the matrix. We will see in a moment that this ensures the absence of a sign problem in the CTQMC simulation. 

We write Eq.(\ref{eq:Z}) in a form suitable for Monte Carlo sampling

\begin{eqnarray}
Z  = Z_{0}\sum_{\mathcal{C}} w(\mathcal{C}) 
\label{eq:ensemble}
\end{eqnarray}
where $\mathcal{C}=\{\mathbf{i}_{1} ,\mathbf{i}_{2}; \tau_{2} \}, \{\mathbf{i}_{3} ,\mathbf{i}_{4}; \tau_{4} \}\ldots \{\mathbf{i}_{2k-1} ,\mathbf{i}_{2k}; \tau_{2k} \}$ denotes a configuration with $k$ vertices. Eq.(\ref{eq:skewsymm}) ensures that the weight $w(\mathcal{C})$~\footnote{The $1/k!$ factor has been canceled by the $k!$ permutations of the vertices.} is always positive~\cite{Huffman:2014fj}

\begin{eqnarray}
w(\mathcal{C}) & =& (-V)^{k} \det(G^{k})
  =   ( -V )^{k} \det (D^{k}) \det({{G}^{k}} D^{k} )\nonumber \\
  & = &  V^{k}  \mathrm{pf} ({G}^{k} D^{k} )^{2} \ge 0, 
\end{eqnarray}
In the second line we have used $\det ( D^{k}) = \prod_{\ell =1}^{k} \eta_{\mathbf{i}_{2\ell-1}}\eta_{\mathbf{i}_{2\ell}}= ( -1 )^{k}$. The absence of a sign problem allows us to simulate fairly large systems at low temperatures to access the quantum critical point. In this paper, we simulate clusters with $L\times L$ unit cells with periodic boundary conditions. The number of sites is $N_{s}=2L^{2}$. Close to the quantum critical point, nonrelativistic corrections are irrelevant and the dynamical critical exponent $z=1$~\cite{Herbut:2009bb}. We thus scale the inverse temperature linearly with the system length $\beta= 4L/3$. Because of the $\beta^{3}$ scaling of the CTQMC algorithm~\cite{Rubtsov:2005iw,Gull:2011jd}, the largest system size $L=15$ considered in this paper is smaller than the one used in the projective auxiliary field QMC studies of the Hubbard model~\cite{Meng:2010gc, Sorella:2012hia, Assaad:2013kg}.


%


To detect the onset of the CDW order, we measure the density-density correlation function   
\begin{eqnarray}
C(R) = \frac{1}{N_{s}N_{R}}\sum_{\mathbf{i}}\sum_{|\mathbf{j-i}|=R}\left \langle \left(\hat{n}_\mathbf{i} -\frac{1}{2}\right) \left(\hat{n}_{\mathbf{j}} -\frac{1}{2}\right) \right \rangle.
\label{eq:nncorr}
\end{eqnarray} 
where $\langle\ldots\rangle= \Tr (e^{-\beta \hat{H}}\ldots)/Z$ denotes the average over the full Hamiltonian Eq.(\ref{eq:Ham}). The second summation in Eq.(\ref{eq:nncorr}) runs over all sites $\mathbf{j}$ (in total $N_{R}$ of them) whose graph distance to the site $\mathbf{i}$ is $R$~\footnote{Strictly speaking, these sites may not be symmetrically related and may have slightly different correlation functions.}. Two sites with even (odd) graph distance have the same (different) parities. The other two important observables are the square and quartic of the CDW order parameter 

\begin{eqnarray}
M_{2}& = & \frac{1}{N_{s}^{2}} \sum_\mathbf{i,j} \eta_\mathbf{i} \eta_\mathbf{j} \left\langle \left(\hat{n}_\mathbf{i} -\frac{1}{2}\right)\left(\hat{n}_\mathbf{j} -\frac{1}{2}\right) 
  \right\rangle, \label{eq:M2} \\
M_{4} & = & \frac{1}{N_{s}^{4}} \sum_\mathbf{i ,j,k ,l } \eta_\mathbf{i} \eta_\mathbf{j} \eta_\mathbf{k} \eta_\mathbf{l}\left \langle \left( \hat{n}_\mathbf{i} -
\frac{1}{2} \right) \left( \hat{n}_\mathbf{j} - \frac{1}{2} \right) \left( \hat{n}_\mathbf{k} -
\frac{1}{2} \right) \left( \hat{n}_\mathbf{l} - \frac{1}{2} \right) \right\rangle. \label{eq:M4}
\end{eqnarray}

The Binder ratio \cite{Binder:1981uk} is calculated as:
\begin{eqnarray}
B & = & \frac{M_{4}}{(M_{2})^{2}}. 
  \label{eq:binderratio}
\end{eqnarray}

%
  
%
%

\begin{figure}[th]
 \centering

\begin{tikzpicture}[]  
  \coordinate (i) at (0,0);
  \coordinate (j) at (6,0);   
  \draw[|-|] (i) node [left] {$0$} -- (j) node [right] {$\beta$};
 
  \coordinate (i) at (1,0.8);
  \coordinate (j) at (1,-0.8);
  \draw [-](i) -- (j) node [below] {};
  \shade[ball color=gray] (i) circle (1mm) node [right] {$\mathbf{i}_{1}$};
  \shade[ball color=gray] (j) circle (1mm) node [right] {$\mathbf{i}_{2}$};
  \draw (1 ,0) circle (0.5mm) node [below right] {$\tau_{2}$};
    
  \coordinate (i) at (1.7,0.8);
  \coordinate (j) at (1.7,-0.8);
  \draw [-](i) -- (j) node [below] {};
  \shade[ball color=gray] (i) circle (1mm) node [right] {$\mathbf{i}_{3}$};
  \shade[ball color=gray] (j) circle (1mm) node [right] {$\mathbf{i}_{4}$};
  \draw (1.7,0) circle (0.5mm) node [below right] {$\tau_{4}$};
   
  \coordinate (i) at (3.1,0.8);
  \coordinate (j) at (3.1,-0.8);
  \draw [dashed, -](i) -- (j) node [below] {};
  \shade[ball color=red] (i) circle (1mm) node [right] {$\mathbf{i}$};
  \shade[ball color=blue] (j) circle (1mm) node [right] {$\mathbf{j}$};
  \draw (3.1 ,0) circle (0.5mm) node [below right] {$\tau$};
  
  \coordinate (i) at (4.8,0.8);
  \coordinate (j) at (4.8,-0.8);
  \draw [-](i) -- (j) node [below] {};
  \shade[ball color=gray] (i) circle (1mm) node [right] {$\mathbf{i}_{5}$};
  \shade[ball color=gray] (j) circle (1mm) node [right] {$\mathbf{i}_{6}$};
  \draw (4.8 ,0) circle (0.5mm) node [below right] {$\tau_{6}$};
  
\end{tikzpicture}
\caption{A example configuration in the worm space. The red and blue dots denotes the worm sites and the gray dots connected by the solid lines denote the vertices. The Monte Carlo updates consist of adding/removing the vertices and the worm and shifting the spatial/time indices of the worm.}
\label{fig:snapshot}
\end{figure}
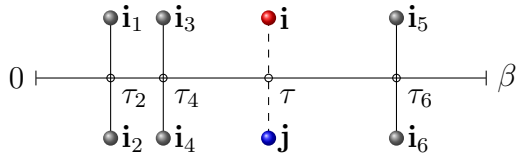

\subsection{Worm Algorithm \label{sec:worm}}
Measuring $M_{2}$ and $M_{4}$ using the conventional approach~\cite{Rubtsov:2005iw} requires explicit loops over the $\mathbf{i},\mathbf{j}, (\mathbf{k},\mathbf{l})$ indices and the measurements will dominate the runtime of Monte Carlo simulations. This is especially inefficient when noticing that each term in Eqs.(\ref{eq:M2}-\ref{eq:M4}) may differ by orders of magnitude. To overcome this difficulty we extend the configuration space and use the worm algorithm~\cite{Prokofev:1998tc, Burovski:2006hv, Gull:2011jd} to sample them efficiently. 

Notice the similarity between the partition function Eq.(\ref{eq:Z}) and the observables Eqs.(\ref{eq:M2}-\ref{eq:M4})

\begin{eqnarray}
 M_{2}&= &
   \frac{1}{\beta N_{s}^{2}} \frac{Z_{0}}{Z}\sum_{k=0}^{\infty} \frac{( -V )^{k}}{k!} \int_{0}^{\beta} \mathrm{d}  \tau_{2} \int_{0}^{\beta} \mathrm{d}  \tau_{4}  \ldots
  \int_{0}^{\beta}  \mathrm{d}  \tau_{2k} \nonumber \times \\ &&  \int_{0}^{\beta} \mathrm{d}  \tau \sum_\mathbf{i,j}\eta_\mathbf{i}\eta_\mathbf{j} \det ( G^{k;\mathbf{ij}\tau}) ,  \label{eq:W2}   \\
M_{4} & = & \frac{1}{\beta N_{s}^{4}} \frac{Z_{0}}{Z}\sum_{k=0}^{\infty} \frac{( -V )^{k}}{k!} \int_{0}^{\beta} \mathrm{d}  \tau_{2} \int_{0}^{\beta} \mathrm{d} \tau_{4}  \ldots
  \int_{0}^{\beta}  \mathrm{d}  \tau_{2k} \nonumber \times \\ &&   \int_{0}^{\beta} \mathrm{d}  \tau \sum_\mathbf{i,j,k,l}\eta_\mathbf{i}\eta_\mathbf{j}\eta_\mathbf{k}\eta_\mathbf{l} \det (G^{k;\mathbf{ijkl}\tau}),
  \label{eq:W4}
\end{eqnarray}
where $G^{k;\mathbf{ij}\tau}$ extends $G^{k}$ to the following $(2k+2)\times(2k+2) $ matrix:
\begin{equation}
\newcommand*{\temp}{\multicolumn{1}{r|}{}} 
G^{k;\mathbf{ij}\tau}=\left(\begin{array}{cccc}
\mathcal{G}^{0}_{\mathbf{i}_{p} \mathbf{i}_{q}}(\tau_{p}-\tau_{q})-\delta_{pq}/2 &\temp &  \mathcal{G}^{0}_{\mathbf{i}_{p} \mathbf{i}}(\tau_{p}-\tau) &  \mathcal{G}^{0}_{\mathbf{i}_{p} \mathbf{j}}(\tau_{p}-\tau)  \\ \cline{1-4} 
\mathcal{G}^{0}_{\mathbf{i} \mathbf{i}_{q}}(\tau-\tau_{q})   &\temp& 0 & \mathcal{G}^{0}_{\mathbf{i} \mathbf{j}}(0^{+})\\
\mathcal{G}^{0}_{\mathbf{j}\mathbf{i}_{q}}(\tau-\tau_{q})  &\temp&\mathcal{G}^{0}_{\mathbf{j} \mathbf{i}}(0^{-})  & 0
\end{array}\right).
\label{eq:Gkijtau}
\end{equation}
Similar to Eq.(\ref{eq:skewsymm}), $G^{k;\mathbf{ij}\tau}$ satisfies the following equation  

\begin{eqnarray}
 (G^{k;\mathbf{ij}\tau} D^{k;\mathbf{ij}})^{T}  & =& - G^{k;\mathbf{ij}\tau} D^{k;\mathbf{ij}} 
\end{eqnarray}
where the diagonal matrix $D^{k;\mathbf{ij}}  =  \diag(\eta_{\mathbf{i}_{1}},\ldots, \eta_{\mathbf{i}_{2k}}, \eta_{\mathbf{i}},\eta_{\mathbf{j}})$ extends $D^{k}$ in a similar manner. Similarly, $G^{k;\mathbf{ijkl}\tau}$ is a $(2k+4)\times(2k+4) $ matrix. We define $W_{2}= \xi_{2}\beta N_{s}^{2}Z M_{2}$,  $W_{4}= \xi_{4}\beta N_{s}^{4}Z M_{4}$ and enlarge the configuration space into 
 
\begin{equation}
Z+  W_{2}+ W_{4} = Z_{0} \sum_{\mathcal{C}} w(\mathcal{C}) . 
\label{eq:wormspace}
\end{equation}
Now the configurations $\mathcal{C}$ may contain a two site worm $\{\mathbf{i} ,\mathbf{j}; \tau \}$ or a four site worm $\{\mathbf{i} ,\mathbf{j}, \mathbf{k}, \mathbf{l}; \tau \}$ in addition to the vertices described in Eqs.(\ref{eq:ensemble}).  By sampling the extended configuration space we can treat the summation over $\mathbf{i},\mathbf{j},\mathbf{k},\mathbf{l}$ in Eq.(\ref{eq:W2}-\ref{eq:W4}) and the summations over the vertices on an equal footing. Here $\xi_{2}$ and $\xi_{4}$ are two positive numbers we can choose freely to balance the configurations in different sectors. We have devised several Monte Carlo updates and describe them in~\ref{sec:updates}. We use the following notation to denote the relative time spend in the each sector~\cite{Burovski:2006hv} 
 
\begin{eqnarray}
  \langle \delta^{Z} \rangle_{\mathrm{MC}} & = & \frac{Z}{Z+ W_{2} + W_{4} }  , \\
  \langle \delta^{W_{2}} \rangle_{\mathrm{MC}} & = & \frac {W_{2}}{Z+ W_{2} + W_{4} }, \\
  \langle \delta^{W_{4}} \rangle_{\mathrm{MC}} & = &\frac {W_{4}}{Z+  W_{2} + W_{4} } . 
  \end{eqnarray}
  
The observables (\ref{eq:M2}-\ref{eq:binderratio}) then are 
\begin{eqnarray}
  M_{2}& = & \frac{1}{\xi_{2}\beta N_{s}^{2}}
  \frac{\langle \delta^{W_{2}} \rangle_{\mathrm{MC}}}{\langle \delta^{Z}\rangle_{\mathrm{MC}}}, \\
    M_{4}& = & \frac{1}{\xi_{4}\beta N_{s}^{4}}
  \frac{\langle \delta^{W_{4}} \rangle_{\mathrm{MC}}} {\langle \delta^{Z}\rangle_{\mathrm{MC}}} ,\\
B & = & \frac{\beta\xi_{2}^{2}}{\xi_{4}} \frac{\langle \delta^{W_{4}} \rangle_{\mathrm{MC}} \langle \delta^{Z} \rangle_{\mathrm{MC}}  }{ (\langle \delta^{W_{2}} \rangle_{\mathrm{MC}} ) ^{2}}.
\end{eqnarray} 
The density correlation function is measured when the configuration is in the $W_{2}$ space and the distance between the two worm sites $\mathbf{i,j}$ is equal to $R$,
\begin{equation}
  C(R) = \frac{1}{\xi_{2}\beta N_{s}N_{R}}
  \frac{\langle \delta^{W_{2}} \delta^{|\mathbf{i-j}|=R}\eta_\mathbf{i}\eta_\mathbf{j} \rangle_{\mathrm{MC}}}{\langle \delta^{Z}
  \rangle_{\mathrm{MC}}}. 
\end{equation} 

The weight of a configuration $\mathcal{C}\in W_{2}$  with worm at $\{\mathbf{i,j};\tau\}$ is 

\begin{eqnarray} 
w(\mathcal{C})  =  ( -V )^{k} \eta_\mathbf{i} \eta_\mathbf{j} \det ( G^{k;\mathbf{ij}\tau}) 
   = V ^{k} \mathrm{pf} (  G^{k;\mathbf{ij}\tau} D^{k;\mathbf{ij}}  )^{2} \ge 0
\end{eqnarray}
where we have used $\det  (D^{k;\mathbf{ij}}) = (\prod_{\ell =1}^{k}
\eta_{\mathbf{i}_{2{\ell}-1}} \eta_{\mathbf{i}_{2\ell}} ) \eta_\mathbf{i} \eta_\mathbf{j}$=$( -1 )^{k}  \eta_\mathbf{i} \eta_\mathbf{j}$. One can similarly show that the weight of $\mathcal{C}\in W_{4}$ sector is positive. Therefore, there is no sign problem in the extended configuration space with worms.

\section{Results\label{sec:results}}

\subsection{Quantum Monte Carlo Results}

\begin{figure}[t]
\centering
\includegraphics[width=9cm]{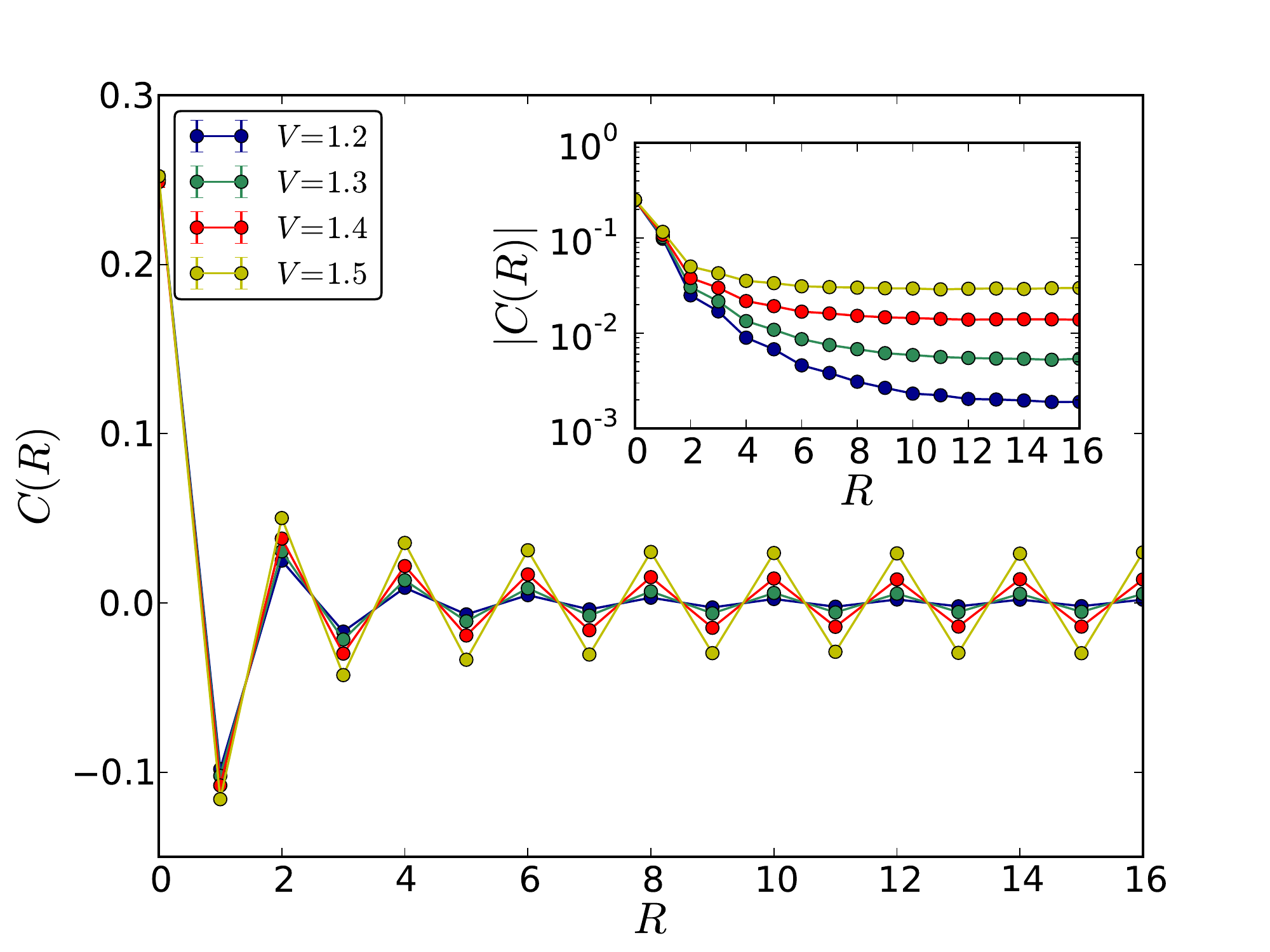}
\caption{The density-density correlations versus distances $R$ on an $L=12$ lattice at $\beta=16$. The inset shows the absolute value of the correlation function on a logarithmic scale. }
\label{fig:nncorr}
\end{figure}

\begin{figure}[t]
\centering
\includegraphics[width=9cm]{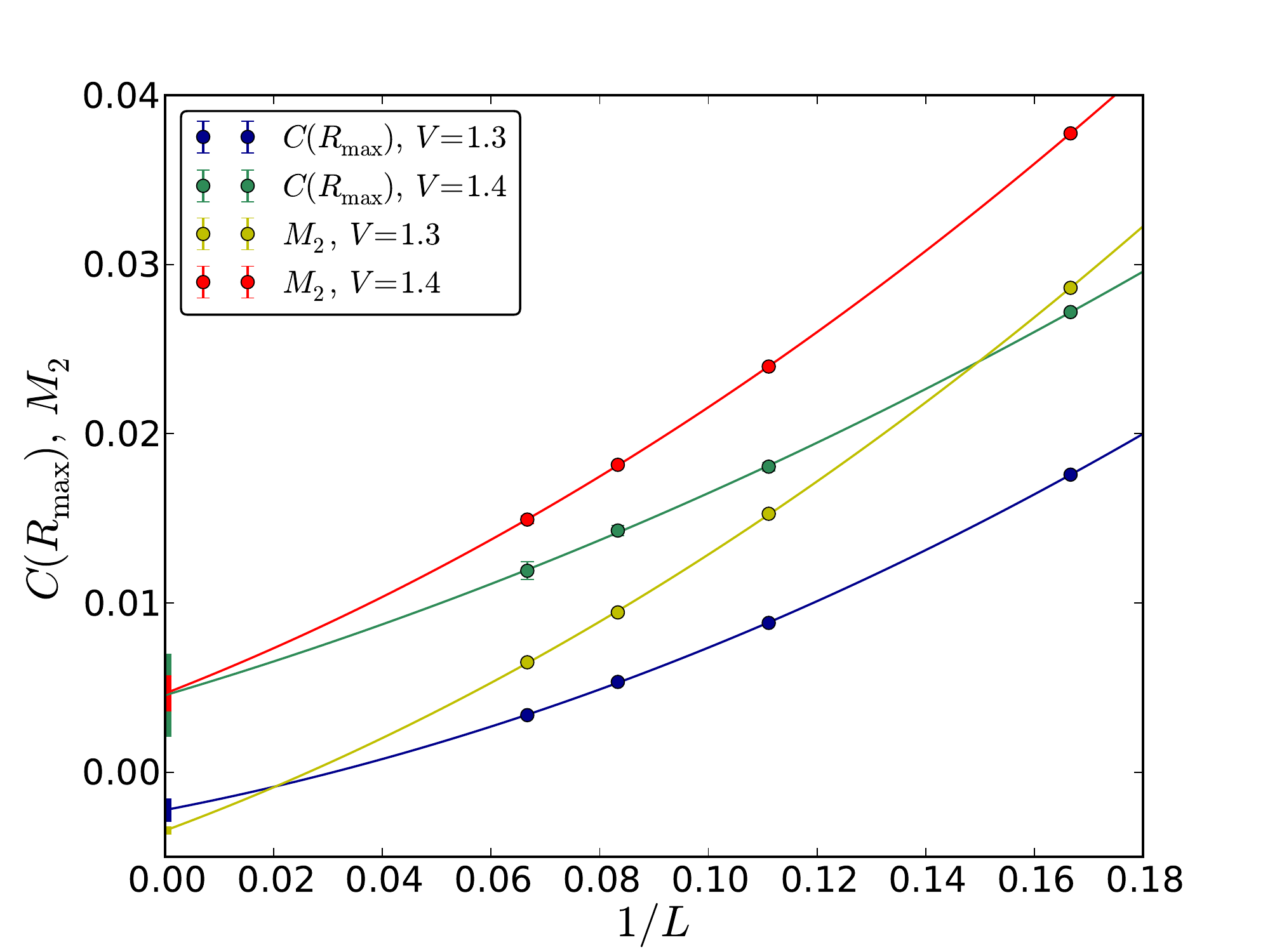}
\caption{Extrapolation of the density correlations at the largest distance $C(R_\mathrm{max})$ and the CDW structure factor $M_{2}$ using $a/L+b/L^{2}$. The error bar of the extrapolation to the thermodynamic limit ($1/L=0$) is evaluated using a jackknife analysis.}
\label{fig:extrapolation}
\end{figure}

Figure~\ref{fig:nncorr} shows the density-density correlations (\ref{eq:nncorr}), which develop a staggered pattern as the interaction strength $V$ increases. The density correlation at the farthest distance $C(R_\mathrm{max})$ and the CDW structure factor $M_{2}$ approach the square of the CDW order parameter as the system size increases. Figure~\ref{fig:extrapolation} shows the extrapolation of $C(R_\mathrm{max})$ and $M_{2}$ to the thermodynamic limit using $a/L+b/L^{2}$. The extrapolation suggests that the quantum critical point lies between $V=1.3$ and $V=1.4$. \footnote{The system sizes up to $L=15$ do not allow us to reliably pin down the critical point based on $1/L$ extrapolation~\cite{Meng:2010gc, Sorella:2012hia}. }

\begin{figure}[th]
\centering
\includegraphics[width=9cm]{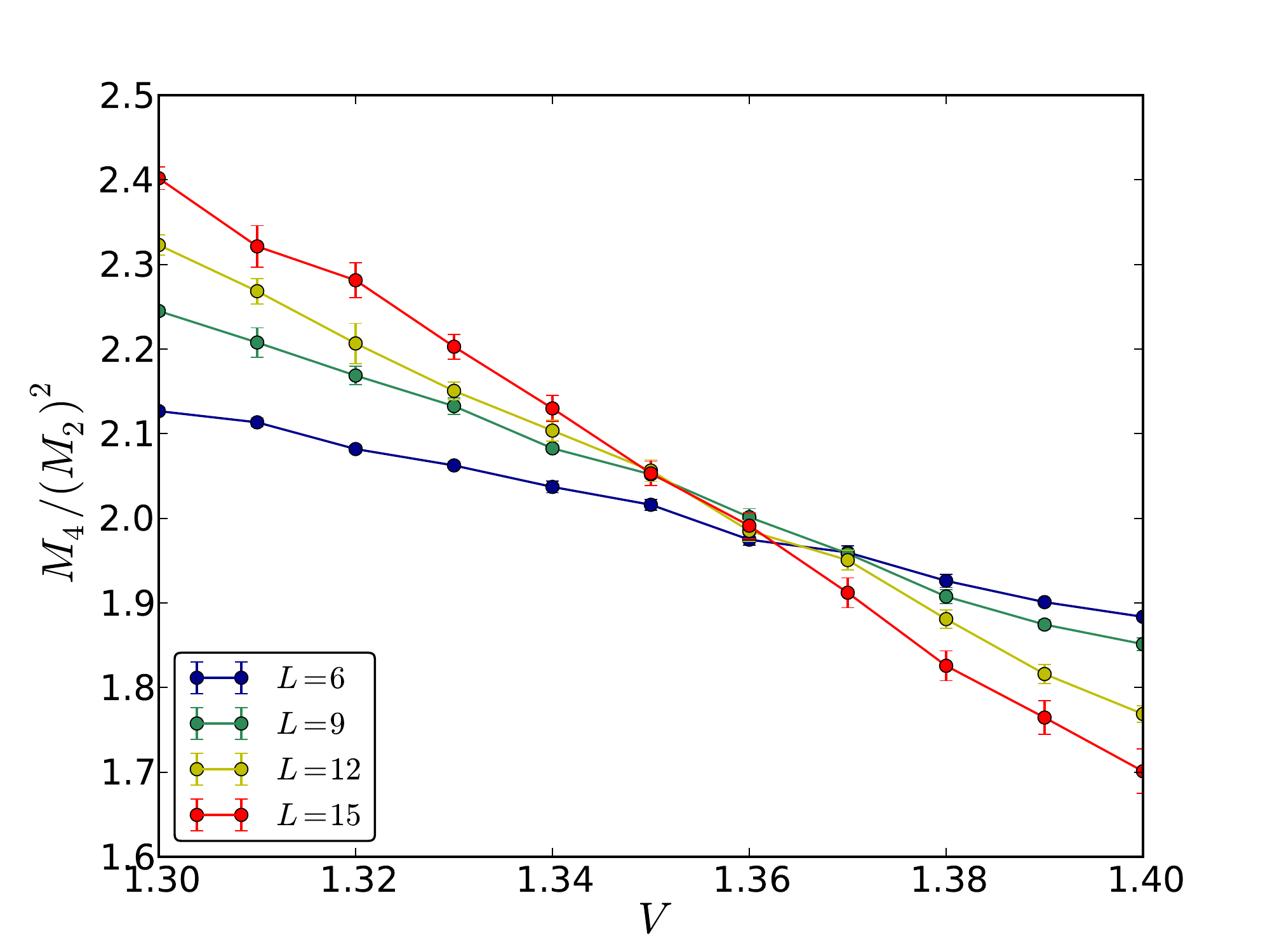}
\caption{The Binder ratio Eq.(\ref{eq:binderratio}) versus $V$ for different system sizes. Lines are linear interpolations of the data. 
}
\label{fig:bindercrossing}
\end{figure}

To better estimate the the critical point we perform a finite size scaling (FSS) analysis based on the scaling ansatz 
\begin{eqnarray}
  M_{2} & = & L ^{-z-\eta} \mathcal{F} ( L^{1/ \nu} (
  V-V_{c} ) ,L^{z} / \beta )  \\
  M_{4} & = & L ^{-2z-2\eta} \mathcal{G}( L^{1/ \nu} ( V-V_{c} ) ,L^{z} / \beta ) \\
    B & = & \mathcal{G}/ \mathcal{F} ^{2}
 \label{eq:fss}
\end{eqnarray}
where $\mathcal{F}$ and $\mathcal{G}$ are universal functions and $\nu,\eta$ are the critical exponents. This scaling ansatz holds close to the critical point. The Binder ratios of different system sizes cross at the transition point. This provides a rough estimate of the transition point $V_{c} \lesssim 1.36$, as shown in Fig.~\ref{fig:bindercrossing}. 


\begin{table}
\caption{The critical point and critical exponents determined using data collapses of $M_{2}$ and $M_{4}$ for all systems sizes ($L=6,9,12,15$) and excluding the smallest one ($L=9,12,15$). The critical exponent $\tilde{\beta}$ (to avoid confusion with the inverse temperature $\beta$) is calculated using $\tilde{\beta} = \frac{\nu}{2}(z+\eta)$. The estimated uncertainty~\cite{Houdayer:2004ii} of the last digit is shown in the bracket. The $\chi^{2}/d.o.f$ listed in the last row shows the quality of the data collapse. 
\label{tab:exponents}} 
\begin{indented}
\centering
\item[]\begin{tabular}{@{}*{5}{l}}
\br                              
 & \centre{2}{$L=6,9,12,15$ }   &  \centre{2}{$L=9,12,15$  } \\ \ns
 & \crule{2} & \crule{2} \\ \ns
  & \centre{1}{$M_{2}$} & \centre{1}{$M_{4}$} &\centre{1}{$M_{2}$} & \centre{1}{$M_{4}$}  \\ \ns
\mr
$V_{c}$         & 1.356(1)   & 1.354(1)   & 1.356(2)  & 1.357(1)   \\
$\nu$           & 0.80(3)    & 0.80(4)    & 0.83(8)   & 0.80(9)  \\
$\eta$          & 0.302(7)   & 0.300(5)   & 0.298(2)   & 0.30(1)  \\
$\tilde{\beta}$          & 0.52(2)   & 0.52(3)    & 0.44(5)   & 0.52(6)  \\ \mr
$\chi^{2}/d.o.f$  & 1.23     &2.05      & 1.4     & 1.61\cr
\br
\end{tabular}
\end{indented}
\end{table}

\begin{figure}[t]
\centering
\includegraphics[width=9cm]{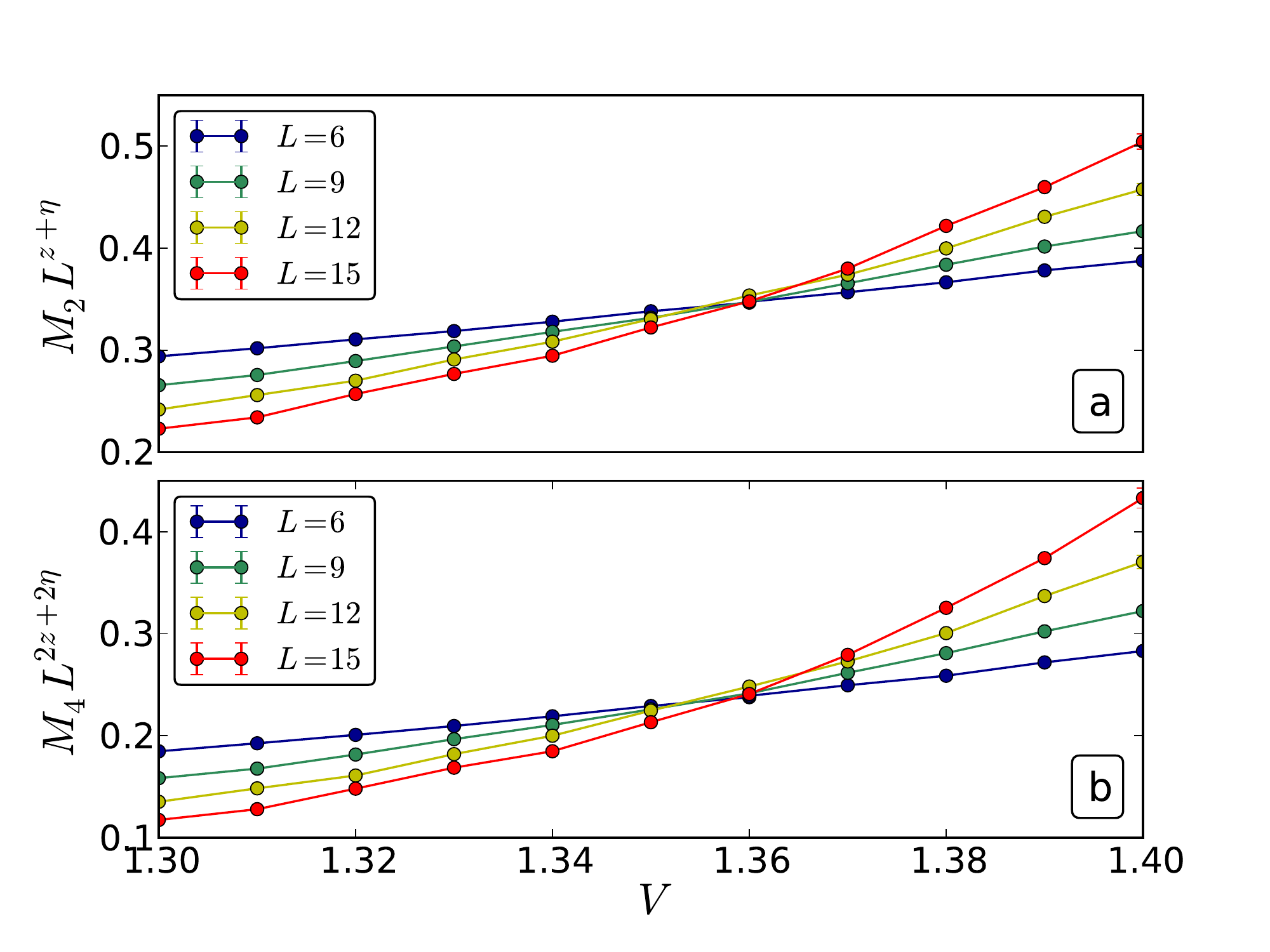}
\includegraphics[width=9cm]{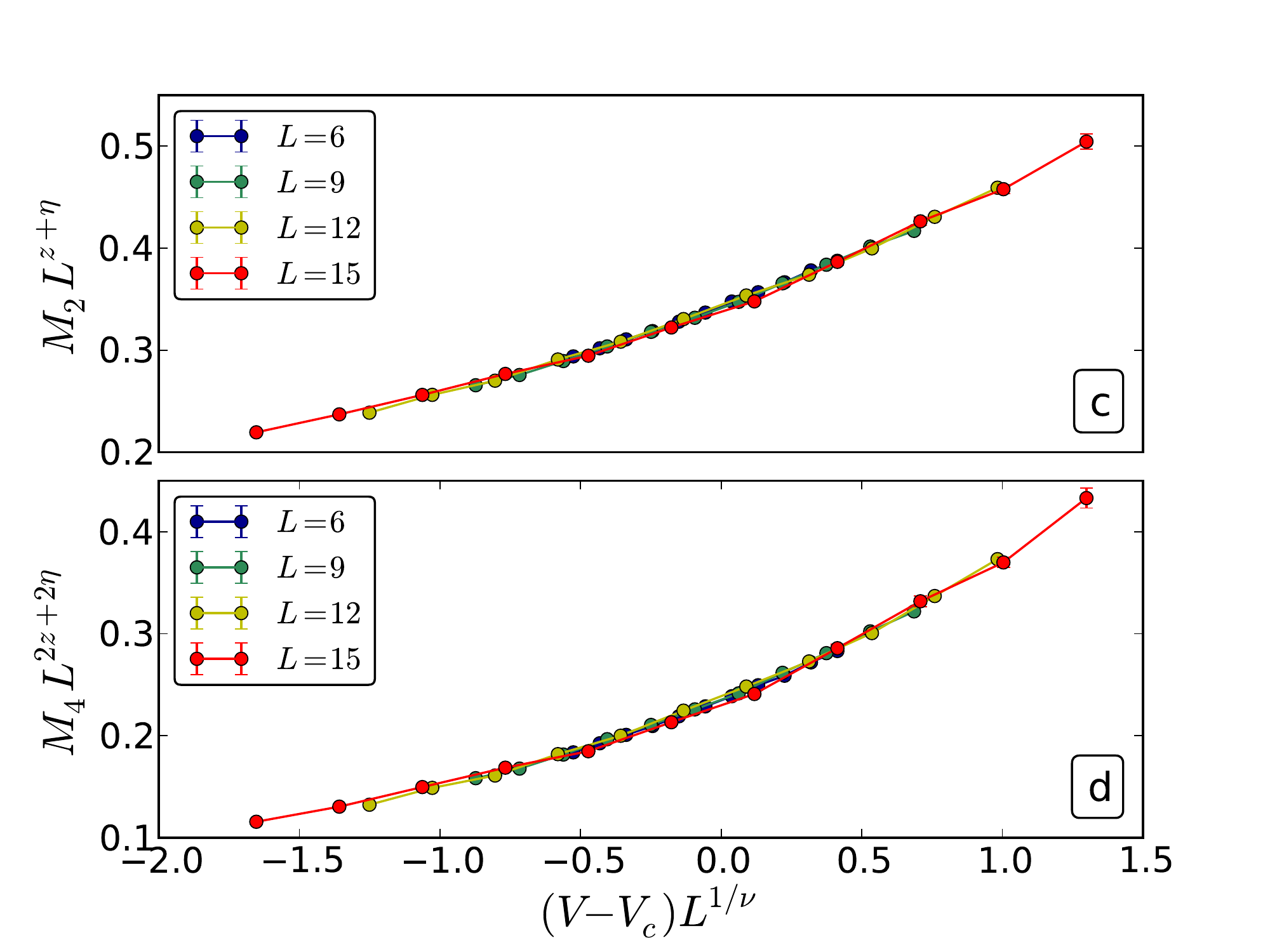}
\caption{(a-b) The scaled $M_{2}$ and $M_{4}$ using $\eta = 0.3$. (c-d) Data collapse using $V_{c}=1.356$ and $\nu = 0.8 $.}
\label{fig:datacollapse}
\end{figure}
 
We next collapse the data of $M_{2}$ and $M_{4}$ to determine the transition point $V_{c}$ and the critical exponents $\nu,\eta$. The results are summarized in Table~\ref{tab:exponents} where we also list the estimates of the order parameter critical exponent $\tilde{\beta}= \frac{\nu}{2}(1+\eta)$, which we will compare with iPEPS results in Sec.\ref{sec:iPEPS}. We have performed the data collapse  using all available system sizes ($L=6,9,12,15$) and excluding the smallest system size ($L=6$). They both give satisfactory data collapse where the $\chi^{2}$ per degree of freedom (d.o.f) is close to one. To visually examine the quality of the data collapse, Fig.~\ref{fig:datacollapse}(a-b) shows the scaled $M_{2}$ and $M_{4}$ using $\eta = 0.3$ where all the curves intersect around $V=1.36$. Further scale the horizontal axis using $V_{c}=1.356$ and $\nu=0.8$ collapse all the data onto a single curve, Fig.\ref{fig:datacollapse}(c-d). 

Our estimation of the correlation length exponent $\nu$ agrees with earlier $\epsilon$-expansion result $\nu=0.797$~\cite{Rosenstein:1993wh} and functional renormalization group results $\nu=0.738\sim0.927$~\cite{Rosa:2001en,Hofling:2002bl}. However, our estimated anomalous dimension $\eta\approx 0.3 $ is smaller than the previous estimates $\eta=0.502\sim0.635$~\cite{Rosenstein:1993wh,Rosa:2001en,Hofling:2002bl}. We have checked that these values of $\eta$ are not consistent with our QMC data. These field theory calculations~\cite{Rosenstein:1993wh,Rosa:2001en,Hofling:2002bl} treated Dirac fermions with the same chiralities but our lattice model contains two Dirac fermions with opposite chirality. This difference might explain the discrepancies with our QMC data. On the other hand, since we observed subleading corrections in the Binder ratio crossing Fig.~\ref{fig:bindercrossing}, further research using larger systems may be needed to determined the critical behavior more accurately. 



 
\subsection{Comparison with iPEPS Results~\label{sec:iPEPS}}

As an independent check of the results we have studied the model (\ref{eq:Ham}) with infinite projected entangled-pair states (iPEPS) - a variational tensor-network ansatz for two-dimensional ground-state wave functions in the thermodynamic limit~\cite{verstraete2004,Verstraete08,jordan2008,corboz2010}. This ansatz is a natural extension of matrix product states (the underlying ansatz of the density-matrix renormalization group method) to two dimensions, and has been previously applied to the same model for attractive interactions~\cite{Corboz12_comment, gu13}. Two-dimensional tensor networks have first been introduced for spin models and later extended to fermionic systems~\cite{Corboz10_fmera, kraus2010, pineda2010,  barthel2009, shi2009, Corboz09_fmera,corboz2010,gu2010}. 
 The iPEPS ansatz consists of a cell of tensors with one tensor per lattice site, which is periodically repeated on the lattice. Each tensor has a physical index of dimension $d$ which carries the local Hilbert space of a lattice site and $z$ auxiliary indices which connect to the $z$ nearest-neighboring tensors. The number of variational parameters (i.e. the accuracy of the ansatz) can be controlled by the bond dimension $D$ of the auxiliary indices, where each tensor contains $dD^z$ variational parameters with $d=2$ and $z=3$ for the present model. For a general introduction to fermionic iPEPS we refer to Ref.~\cite{corboz2010}.  

We simulate the honeycomb model (\ref{eq:Ham}) by mapping it onto a brick-wall square lattice, as done in Ref.~\cite{Corboz12_su4}. The variational parameters of the iPEPS ansatz are optimized by performing an imaginary time evolution using a second order Trotter-Suzuki decomposition and the full-update scheme for the truncation of a bond index (see Ref.~\cite{corboz2010} for details). To evaluate the iPEPS wave function (e.g. for the computation of expectation values) we use a variant of the corner-transfer-matrix method~\cite{nishino1996, orus2009-1} described in Refs.~\cite{corboz2011,corboz2014}. The U(1) symmetry of the present model is exploited~\cite{singh2010,bauer2011} to increase the efficiency of the simulations.

\begin{figure}[t]
\centering
\includegraphics[width=12cm]{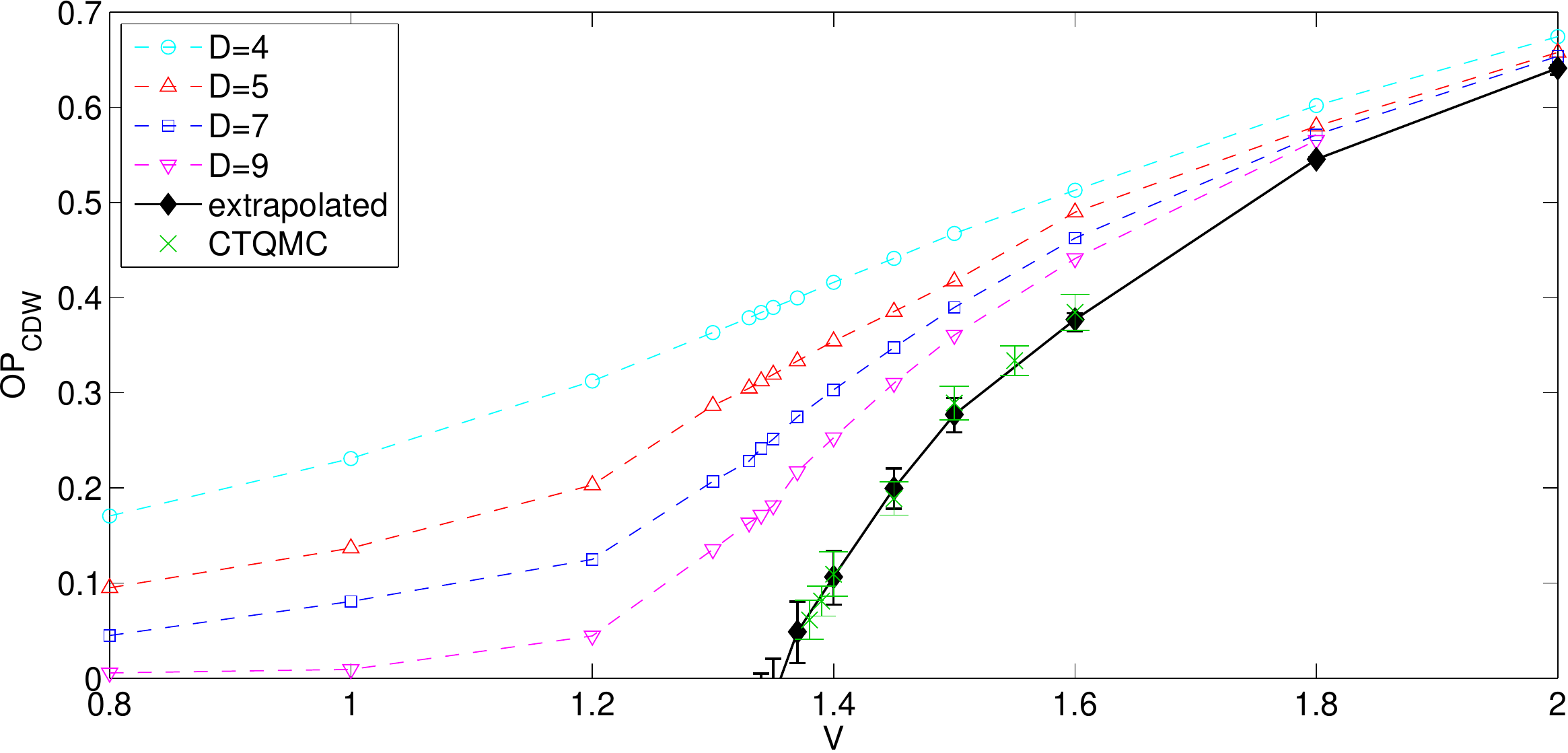}
\caption{Results for the CDW order parameter as a function of repulsion strength $V$ obtained with iPEPS, compared to the extrapolated CTQMC data.}
\label{fig:iPEPS}
\end{figure}

Since iPEPS represent a wave function in the thermodynamic limit, symmetries of the Hamiltonian can be spontaneously broken, and the order can be measured by a local order parameter. In Fig.~\ref{fig:iPEPS} the iPEPS results for the CDW order parameter $OP_{CDW}=\left| \langle \hat n_A - \hat n_B \rangle \right|$ as a function of $V$ is shown, where $\hat n_A$ and  $\hat n_B$ correspond to the particle density on sublattices A and B, respectively. Since iPEPS is an ansatz in the thermodynamic limit, there are no finite size effects. However, the finite bond dimension $D$ has a similar effect on the order parameter as a finite size system, i.e. there is no sharp transition but the order parameter is overestimated around the critical coupling $V_c$. To obtain an estimate of the  order parameter in the infinite $D$ limit we extrapolate the data linearly in $1/D$, shown by the black diamonds in Fig.~\ref{fig:iPEPS}. The error bar indicates the range of extrapolated values by taking into account different sets of data points. Although the analytical dependence of the order parameter on $D$ is not known, empirically, one can get a reasonable estimate by such type of extrapolations (see e.g. Ref~\cite{Corboz13_shastry}). Based on these extrapolations of the iPEPS data up to $D=9$ we obtain a value of the critical coupling of $V_c=1.36(3)$, in agreement with the CTQMC result.

The green crosses in Fig.~\ref{fig:iPEPS} show the CTQMC data for the order parameter in the thermodynamic limit, computed as $OP_{CDW}= \lim_{L\rightarrow \infty} 2\sqrt{M_2(L)}$, which is in agreement with the iPEPS data. Similar results are obtained by estimating the order parameter from $C(R_\mathrm{max})$, i.e. $OP_{CDW}= \lim_{L\rightarrow \infty} 2\sqrt{C(R_\mathrm{max})}$. The extrapolation of QMC data close to the critical point is more difficult because the intersections at $1/L=0$ may become negative. 

We also tried to extract the critical exponent $\tilde{\beta}$ by fitting the extrapolated iPEPS data to $k (V-V_c)^{\tilde{\beta}}$ in the range $[V_c,1.6]$. However, due to the error bars and sensitivity of the exponent on the fitting range we can only give a  crude estimate of $\tilde{\beta}=0.7(15)$, which is somewhat larger than the CTQMC result $\tilde{\beta}=0.52(6)$, but both are smaller than the mean-field result $\tilde{\beta}_{MF}=1$~\cite{Sorella:1992wd} and consistent with the concave shape of the order parameter versus $V$ curve.

\section{Conclusion and Outlook \label{sec:outlook}}

We presented a sign problem free CTQMC study of the Dirac semi-metal to charge-density-wave transition on the honeycomb lattice and compare it with theory and iPEPS results. Our main results about the transition point and the critical exponents are summarized in the Table~\ref{tab:exponents}. The present study uses the static density-density correlations as the diagnosis tool for the quantum critical point, it is however interesting to further study the transport and entanglement properties across the phase transition.  
Future studies may map out the finite temperature phase diagram and especially the crossover~\cite{CardyBook} from the Gross-Neveu to the 2D Ising universality class. The CDW transition of the spinful Dirac fermions~\cite{Honerkamp:2008iw,PhysRevB.89.205128} can also be studied using a similar method. Generalization of the model to include hopping and interactions beyond the nearest neighbors may allow us to address the intriguing question about the emergence~\cite{Raghu:2008kra} and stability of the topological insulating states~\cite{Varney:2010eja,Varney:2011jx} in the presence of interactions. 

\ack
The authors thank Igor Herbut, Kun Yang, Ziyang Meng, Su-Peng Kou, Hong Yao, Jakub Imri\v ska,  Jan Gukelberger and Hiroshi Shinaoka for useful discussions. Simulations were performed on the M\"{o}nch cluster of
Platform for Advanced Scientific Computing (PASC), the Brutus cluster at ETH Zurich, and the ``Monte Rosa'' Cray XE6 at the Swiss
National Supercomputing Centre (CSCS). We have used ALPS libraries~\cite{BBauer:2011tz} for Monte Carlo
simulations and data analysis. This work was supported by ERC Advanced Grant SIMCOFE.

\clearpage 

\appendix


\section{Monte Carlo Updates\label{sec:updates}}

A Monte Carlo update consists of proposing a move from a configuration $\mathcal{C}$ to a new configuration $\mathcal{C}^{\prime}$ with a priori probability $A(\mathcal{C}\rightarrow \mathcal{C}^{\prime})$. The acceptance probability $R(\mathcal{C}\rightarrow \mathcal{C}^{\prime})$ satisfies the detailed balance condition 

\begin{equation}
 R(\mathcal{C}\rightarrow \mathcal{C}^{\prime}) w(\mathcal{C})A(\mathcal{C}\rightarrow \mathcal{C}^{\prime}) = R(\mathcal{C}^{\prime}\rightarrow \mathcal{C})w(\mathcal{C}^{\prime})A(\mathcal{C}^{\prime}\rightarrow \mathcal{C}) 
 \label{eq:detailedbalance}
\end{equation}
The Metropolis-Hasting solution of the detailed balance equation Eq.(\ref{eq:detailedbalance}) is \begin{equation}
R(\mathcal{C}\rightarrow \mathcal{C}^{\prime}) = \min\left\{1, \frac{w(\mathcal{C}^{\prime})A(\mathcal{C}^{\prime}\rightarrow \mathcal{C})  }{w(\mathcal{C})A(\mathcal{C}\rightarrow \mathcal{C}^{\prime})}\right\} 
\end{equation}

\begin{figure}
\centering
\begin{tikzpicture}[->,>=stealth',shorten >=1pt,auto,
  thick,main node/.style={circle,fill=blue!20,draw,font=\sffamily\large\bfseries}]  

  \node[main node]         (Z)  [above =3.8cm]  {$Z$};
  \node[main node]         (W2) [left =2.6cm]  {$W_2$};
  \node[main node]         (W4) [right= 2.6cm] {$W_4$};

 \path[every node/.style={font=\sffamily\footnotesize}]
      (Z)   edge []  node[align=left] {Eq.(\ref{eq:ZtoW2})\\Eq.(\ref{eq:open})} (W2)
            edge [bend left]  node {Eq.(\ref{eq:ZtoW4})} (W4)
            edge [loop above] node {Eq.(\ref{eq:vertexadd}-\ref{eq:vertexremove})} (Z) 
       (W2) edge [bend left]  node[align=left] {Eq.(\ref{eq:W2toZ})\\Eq.(\ref{eq:close})} (Z)
            edge []           node {Eq.(\ref{eq:W2toW4})} (W4)
            edge [loop left]  node[align=left]  {Eq.(\ref{eq:vertexadd}-\ref{eq:vertexremove})\\Eq.(\ref{eq:wormshift})}  (W2)
       (W4) edge []  node {Eq.(\ref{eq:W4toZ})} (Z)
            edge [bend left]  node {Eq.(\ref{eq:W4toW2})} (W2)
            edge [loop right] node[align=left] {Eq.(\ref{eq:vertexadd}-\ref{eq:vertexremove})\\Eq.(\ref{eq:wormshift})}  (W4);
\end{tikzpicture}
\caption{The configuration space and the MC updates.} 
\label{fig:configurations}
\end{figure}
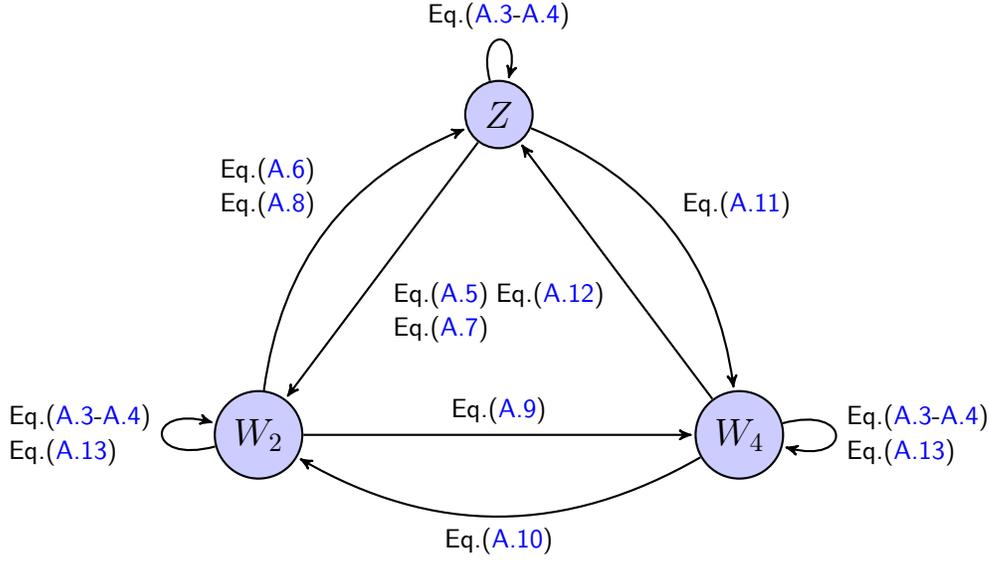

There are three classes of configurations shown in Fig.~\ref{fig:configurations}. We devised several updates to sample the configuration space. Most updates are in complementary pairs. Within each pair one can still fine tune the propose probability to enhance the acceptance rate. 

\subsection{Vertex add/remove}
We add $n$ vertices to a configuration with $k$ vertices. The acceptance ratio is 

\begin{eqnarray}
  R_{\mathrm{add}} & = &\frac{(-V)^{n}\det (G^{k+n}) /{k+n \choose n}}  {\det
  ( G^{k} ) n!(\frac{1}{\beta N_{b}})^{n}}    = (-\beta V N_{b})^{n}\frac{k!}{(k+n)!} \frac{\det (G^{k+n} )}{\det( G^{k} )}  \label{eq:vertexadd}  
\end{eqnarray}
where $N_{b}=3L^{2}$ is the number of bonds of the honeycomb lattice. The move Eq.(\ref{eq:vertexadd}) is balanced by removing $n$ vertices from a $k$-vertices configuration with the acceptance probability
\begin{eqnarray}
  R_{\mathrm{remove}} & = & \frac{1}{(-\beta V N_{b})^{n}} \frac{k!}{(k-n)!}\frac{\det ( G^{k-n} )}{\det
  ( G^{k} )} 
 \label{eq:vertexremove} 
\end{eqnarray}
Replace $G^{k}$ by $G^{k;\mathbf{ij(kl)}\tau}$ one gets the formulas for adding/removing vertices in the worm space.

\subsection{Worm creation/destruction}

\subsubsection{$Z\leftrightarrow W_{2}$}
From the partition function sector we create a worm at $\{\mathbf{i},\mathbf{j};\tau \}$. The corresponding new matrix is $G^{k;\mathbf{ij}\tau}$. To improve the acceptance rate, we select the site $\mathbf{j}$ in the neighborhood (containing $m$ sites) of a randomly chosen site $\mathbf{i}$. The acceptance ratio is 

\begin{eqnarray}
  R_{\mathrm{create}} & = & \eta_\mathbf{i}\eta_\mathbf{j} \xi_{2} N_{s}m \beta \frac{\det (G^{k;\mathbf{ij}\tau} )}{\det ( G^{k} )} \label{eq:ZtoW2} \\
  R_{\mathrm{destroy}} & = &\eta_\mathbf{i}\eta_\mathbf{j}  \frac{1}{\xi_{2} N_{s}m \beta} \frac{\det ( G^{k}
  )}{\det ( G^{k;\mathbf{ij}\tau} )} 
 \label{eq:W2toZ}   
\end{eqnarray}

A cheaper way to go between the partition function and the $W_{2}$ space is to randomly select a vertex and interpret it as a worm. We call this process an open update and the reverse process a close update. These two updates change the perturbation order by one.  However they are cheaper than creating/destroying worms Eq.(\ref{eq:ZtoW2}-\ref{eq:W2toZ}) because no matrix operation is involved. The acceptance ratios are 

\begin{eqnarray}
  R_{\mathrm{open}} & = & 2\xi_{2} \frac{k}{V}   \label{eq:open} \\
  R_{\mathrm{close}} & = & \frac{1}{2\xi_{2}} \frac{V}{(k+1)}    \label{eq:close} 
\end{eqnarray}
The factor $2$ accounts for the fact that $\{\mathbf{i},\mathbf{j};\tau\}$ and $\{\mathbf{j},\mathbf{i};\tau\}$ are counted as two distinct worm configurations.

\subsubsection{$W_{2}\leftrightarrow W_{4}$}

In the $W_{2}$ sector we insert another worm at $\{\mathbf{k},\mathbf{l};\tau \}$ choosing a random site $\mathbf{k}$ and a nearly site $\mathbf{l}$ (out of $m$ sites). The time $\tau$ is the same as the imaginary time of the existing worm $\{\mathbf{i},\mathbf{j};\tau \}$. Acceptance ratios are  

\begin{eqnarray}
  R_{\mathrm{create}} & = & \eta_\mathbf{k}\eta_\mathbf{l}\frac{ N_{s}m \xi_{4}}{\xi_{2}} \frac{\det ( G^{k;\mathbf{ijkl}\tau})}{\det ( G^{k;\mathbf{ij}\tau})}  \label{eq:W2toW4}  \\
  R_{\mathrm{destroy}} & = & \eta_\mathbf{k}\eta_\mathbf{l}\frac{\xi_{2}}{ N_{s}m \xi_{4}} \frac{\det ( G^{k;\mathbf{ij}\tau})}{\det ( G^{k;\mathbf{ijkl}\tau})} 
   \label{eq:W4toW2} 
\end{eqnarray}

\subsubsection{$Z\leftrightarrow W_{4}$}

We create a worm at $\{\mathbf{i},\mathbf{j},\mathbf{k},\mathbf{l};\tau \}$ in the partition function sector. To improve the acceptation ratio, we choose the sites $\mathbf{j,k,l}$ in the neighborhood of a randomly chosen site $\mathbf{i}$. The ratios are

\begin{eqnarray}
  R_{\mathrm{create}} & = & \eta_\mathbf{i}\eta_\mathbf{j}\eta_\mathbf{k}\eta_\mathbf{l} N_{s}m^{3} \beta \xi_{4} \frac{\det ( G^{k;\mathbf{ijkl}\tau})}{\det (
  G^{k})}    \label{eq:ZtoW4} \\
  R_{\mathrm{destroy}} & = & \eta_\mathbf{i}\eta_\mathbf{j}\eta_\mathbf{k}\eta_\mathbf{l}\frac{ 1}{ N_{s}m^{3} \beta \xi_{4} } \frac{\det ( G^{k})}{\det ( G^{k;\mathbf{ijkl}\tau}) } 
  \label{eq:W4toZ}
\end{eqnarray}

\subsection{Worm shift}
We shift the worm to a new space-time point. To enhance the acceptance probability, we randomly choose one site in the worm and shift it to one of its neighbors. The imaginary time $\tau$ is updated to $\tau^{\prime}$ by randomly by adding a random number in the range of $[-0.05\beta, 0.05\beta)$. The matrix is updated to $ G^{k;\mathbf{ij}^{\prime}\tau^{\prime}}$ and the acceptance probability is 

\begin{eqnarray}
  R_{\mathrm{shift}} & = & \eta_\mathbf{j}\eta_{\mathbf{j}^{\prime}}\frac{\det ( G^{k;\mathbf{i}\mathbf{j}^{\prime}\tau^{\prime}} )}{\det ( G^{k;\mathbf{ij}\tau} )}. 
  \label{eq:wormshift}
\end{eqnarray}
This process is self-balanced. The acceptance rate of worm shift in the $W_{4}$ space has a similar expression.


\section{Monte Carlo results on the $\pi$-flux lattice \label{sec:piflux}}

\begin{figure}[h]
\centering
\includegraphics[width=9cm]{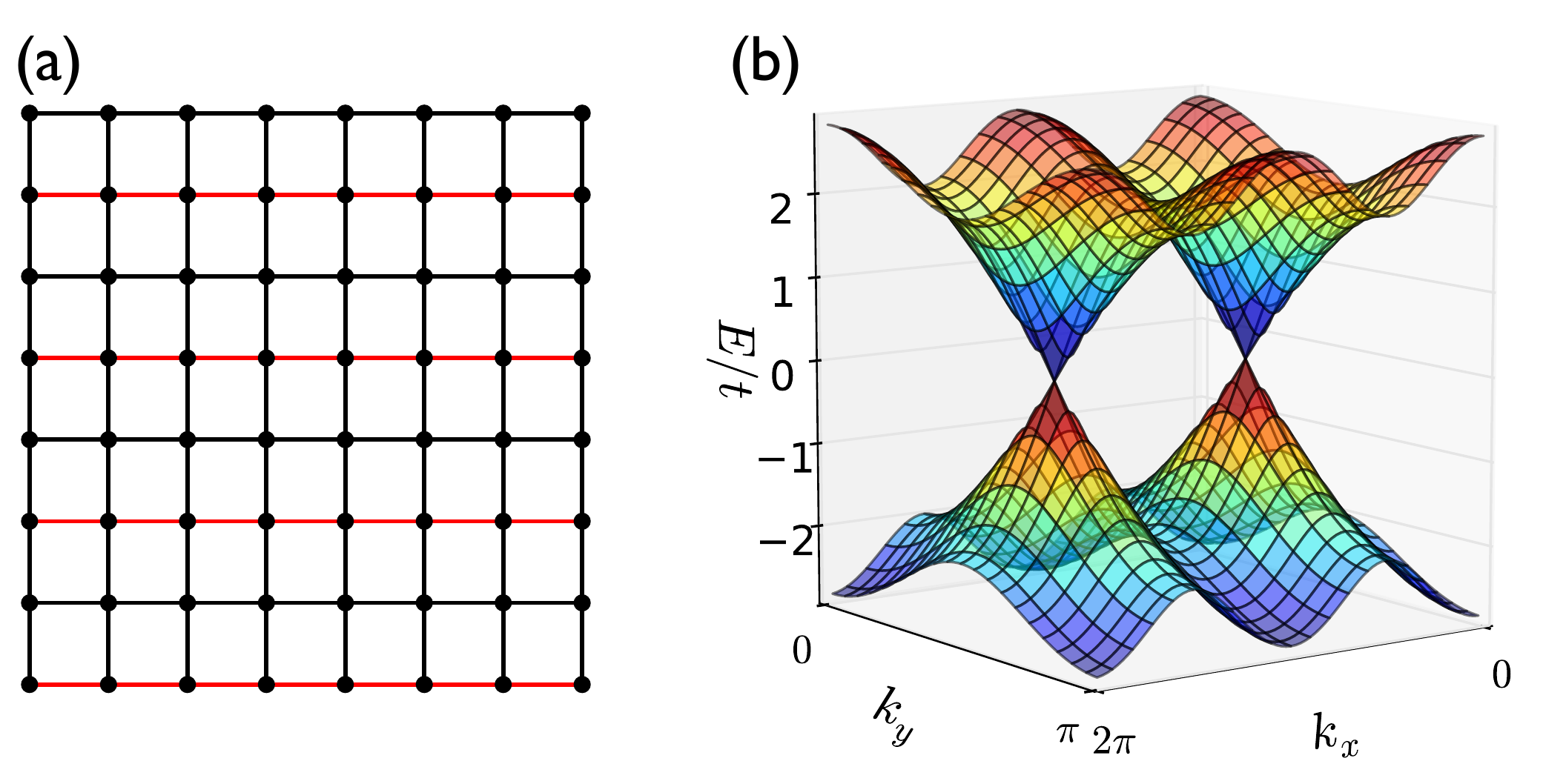}
\caption{(a) A $L=8$ square lattice with $\pi$-flux inserted to each plaquette. The hopping amplitude of the red bonds are $t=-1$, while on the black bonds $t=1$. (b) The noninteracting band structure of the $\pi$-flux lattice. }
\label{fig:pifluxlattice}
\end{figure}

To further confirm the critical exponent found in the main text, we simulated the model Eq.(\ref{eq:Ham})  on a square lattice with $\pi$-flux inserted in each plaquette, Fig.\ref{fig:pifluxlattice}(a). The lattice also features two Dirac points in the Brillouin zone, Fig.\ref{fig:pifluxlattice}(b). The Dirac semimetal to CDW transition should belong to the same universality class in the honeycomb lattice. In the simulation we use the Landau gauge for the flux and choose system sizes $L$ to be divisible by $4$. The inverse temperature scales linearly with length $\beta = L$.

\begin{figure}[th]
\centering
\includegraphics[width=9cm]{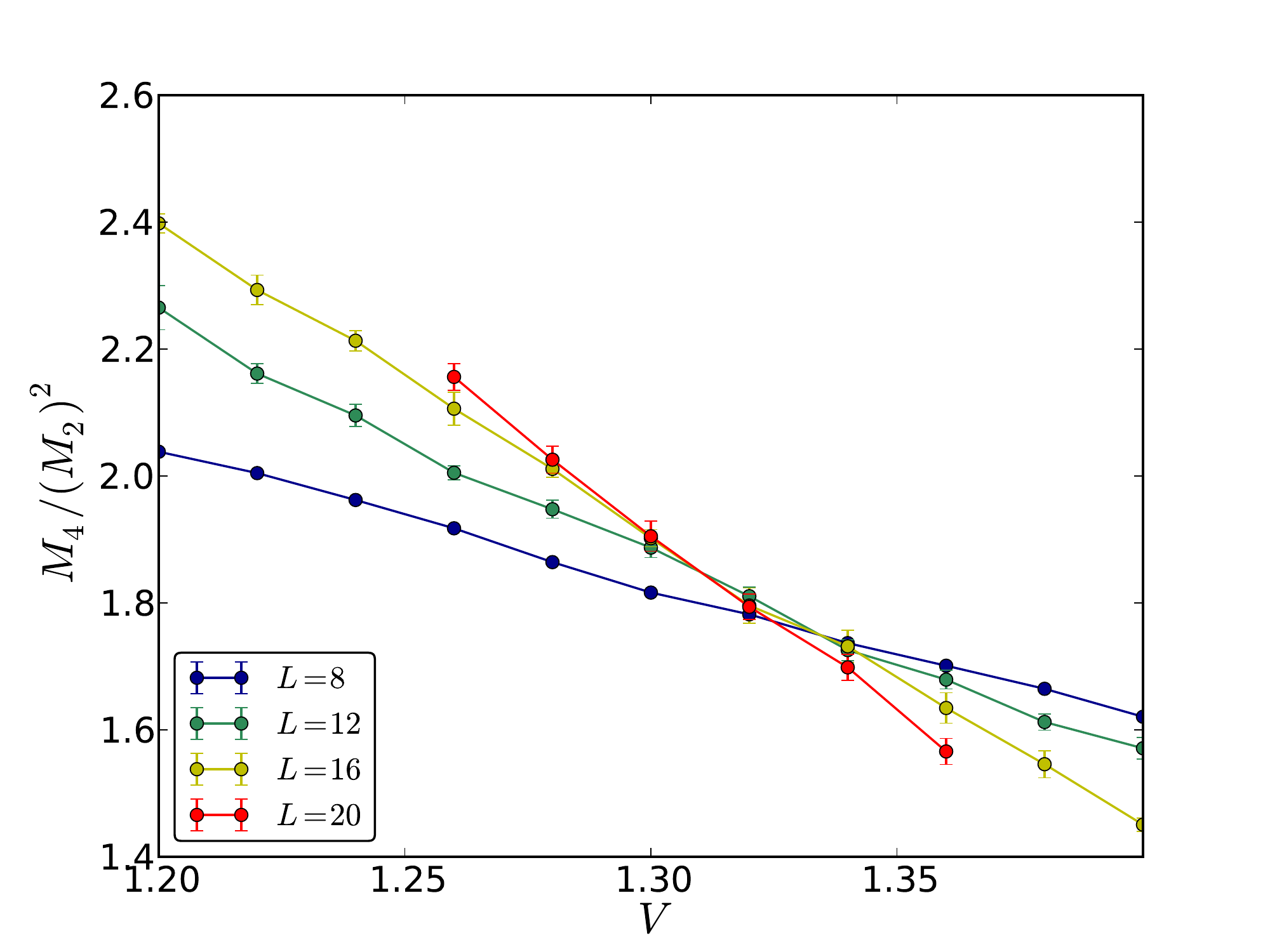}
\caption{The Binder ratio Eq.(\ref{eq:binderratio}) versus $V$ for different system sizes of the $\pi$-flux lattice. Lines are linear interpolations of the data. 
}
\label{fig:piflux_binderratio}
\end{figure}

\begin{figure}[h]
\centering
\includegraphics[width=9cm]{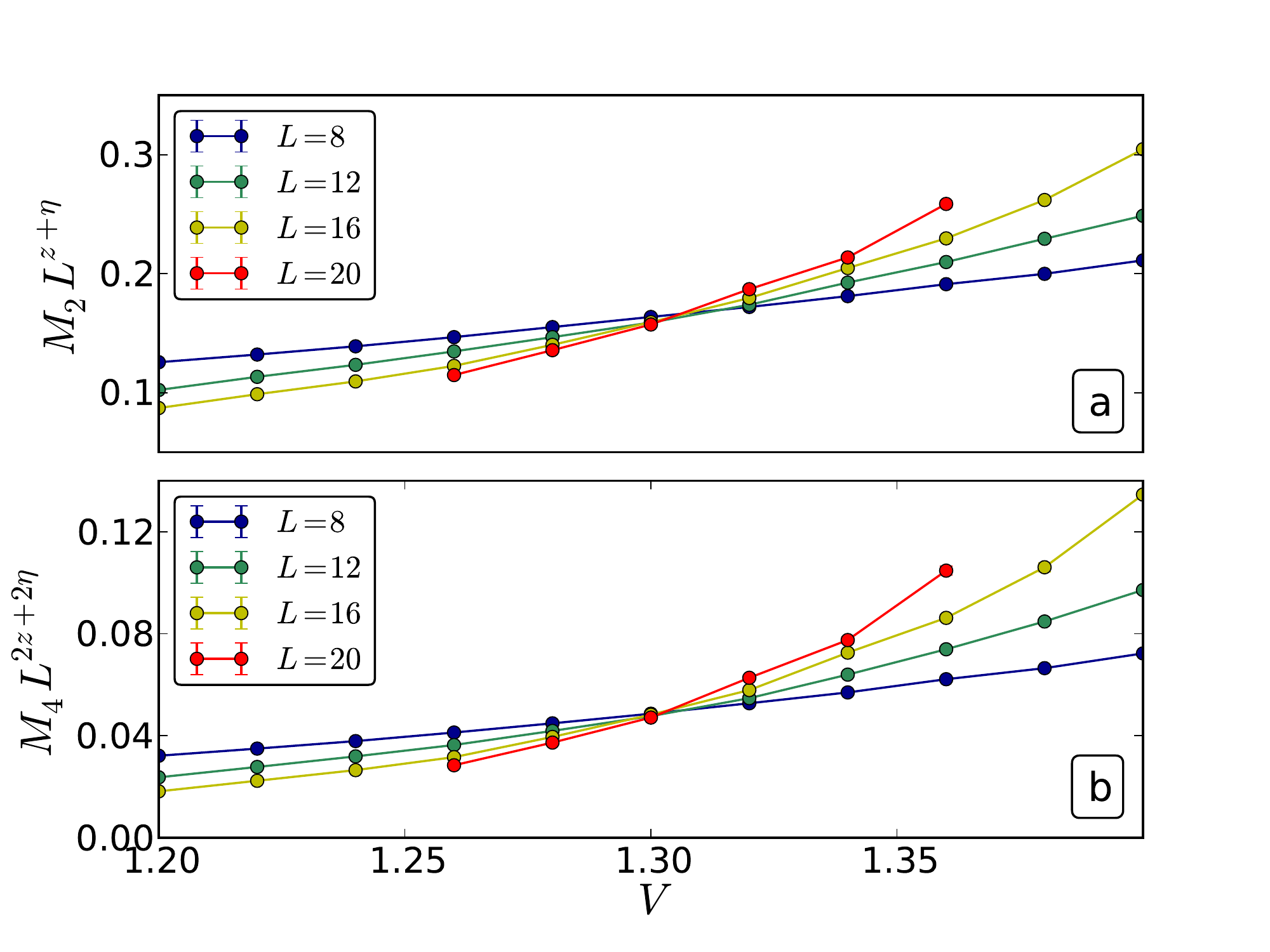}
\includegraphics[width=9cm]{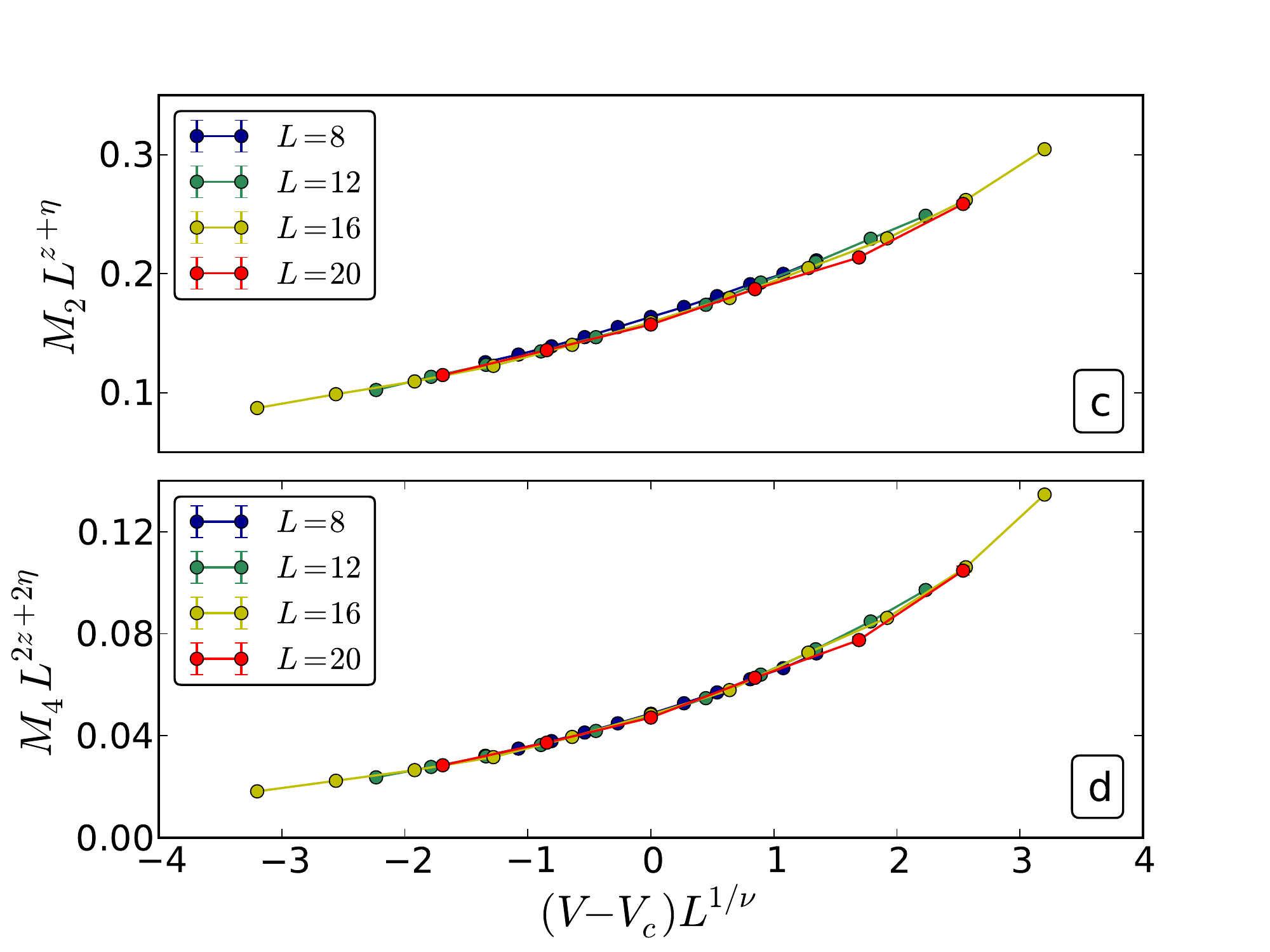}
\caption{(a-b) The scaled $M_{2}$ and $M_{4}$ using $\eta = 0.3$. (c-d) Data collapse using $V_{c}=1.3$ and $\nu = 0.8 $.}
\label{fig:piflux_datacollapse}
\end{figure}

Fig.~\ref{fig:piflux_binderratio} shows the Binder ratios, from which we infer the transition point $V_{c}\approx1.3$. A data collapse analysis of $M_{2}$ (Figure~\ref{fig:piflux_datacollapse})  gives $V_{c}=1.304(2)$, $\nu=0.80(6)$ and $\eta=0.318(8)$. These results indicate that the critical exponents we found for the honeycomb lattice (Table~\ref{tab:exponents}) are universal.

\clearpage

\section*{References}
\bibliography{spinlessCTQMC,refs}
\bibliographystyle{unsrt}

\end{document}